\documentstyle[12pt]{article}
\begin{document}
\begin{titlepage}

\begin{center} \Large {\bf Theoretical Physics Institute}\\ {\bf
University of Minnesota}
\end{center}
\vspace*{.3cm}
\begin{flushright} TPI-MINN-95/18-T \\ UMN-TH-1350-95 \\ OUTP-95-25P \\
hep-th/9507170\\ (revised)
\end{flushright}
\vspace{.3cm}
\begin{center} \Large {\bf  Matching Conditions and  Duality
 in  ${\bf N=1}$ SUSY Gauge Theories in the Conformal Window}
\end{center}
\vspace*{.3cm}
\begin{center} {\large Ian I. Kogan$^{a,b}$, Mikhail Shifman$^b$, and
Arkady  Vainshtein$^{b,c}$} \\
\vspace{0.4cm} {\normalsize
$^a$ {\it Theoretical Physics, 1  Keble Road,
 Oxford, OX1 3NP , UK \footnote{permanent address} \\
$^b$ Theoretical Physics Institute, Univ. of Minnesota, Minneapolis, MN
55455, USA  } \\
$^c$ {\it Budker Institute of Nuclear Physics, Novosibirsk 630090,
Russia}}\\

\vspace*{2cm}

{\Large{\bf Abstract}}
\end{center}

\vspace*{.2cm} We  discuss duality in $N=1$ SUSY gauge theories in
Seiberg's conformal window, $(3N_c/2)<N_f<3N_c$. The 't Hooft
consistency conditions -- the basic tool for establishing the infrared
duality --  are considered taking into account higher order $\alpha$
corrections.  The conserved (anomaly free) $R$ current is built to all
orders in $\alpha$. Although this current contains all orders in
$\alpha$ the 't Hooft consistency conditions for this current are  shown
to be one-loop. This observation thus justifies Seiberg's matching
procedure. We also briefly discuss the inequivalence of the
``electric" and ``magnetic" theories at short distances.

\end{titlepage}

\section{Introduction} In this paper  we   discuss the infrared
duality between different (``electric" and ``magnetic")  $N=1$  SUSY
gauge models observed in  Ref.
\cite{S2}. Supersymmetric (SUSY)  gauge theories are unique  examples of
non-trivial four-dimensional theories where some dynamical aspects  are
exactly tractable. The first results of this type -- calculation of the
gluino condensate and the Gell-Mann-Low function -- were obtained  in
the early eighties \cite{NSVZ1,NSVZ2}. The interest to the miraculous
features of the supersymmetric theories was revived after the recent
discovery \cite{S2}, \cite{S1}-\cite{IS} of a rich spectrum of various
dynamical scenarios that may be realized with a special choice of the
matter sector (for a review see Ref. \cite{Nathan}). The basic tools in
unraveling these scenarios are:
\begin{itemize}
\item  instanton-generated  superpotentials which may or may not lift
degeneracies along  classically flat directions \cite{ADS};
\item  the NSVZ $\beta$ functions \cite{NSVZ1,NSVZ2};
\item   the property of holomorphy in certain parameters
\cite{Shif1,Shif2,S3};
\item  various  general symmetry properties, i.e. the superconformal
invariance at  the infrared fixed points and its consequences \cite{S2}.
\end{itemize}
  A beautiful phenomenon revealed in this way is the existence of a
generalized ``electric-magnetic" duality in $N=2$ \cite{SE} and some
versions of $N=1$ theories \cite{S2}.

  In Ref. \cite{S2} it was argued that $SU(N_c)$ and $SU(N_f-N_c)$
gauge theories with  $N_f$ flavors (and a specific Yukawa interaction
in the ``magnetic" theory) flow to one and the same limit in the
infrared asymptotics. If $3N_c/2 < N_f < 3N_c$ both theories are
conformal -- this is the so called conformal window.  In other words,
for these values of $N_f$ the Gell-Mann-Low functions of both theories
vanish  at critical values of the coupling constants. In particular,
for the  ``electric"
$SU(N_c)$  gauge theory with $N_{f}$ flavors in the fundamental
representation  the $\beta$ function corresponding to the gauge
coupling has the following form \cite{NSVZ1,NSVZ2}:
\begin{eqnarray}
\beta (\alpha ) = -\frac{\alpha^2}{2\pi}\,\,  \frac{3N_c-
N_{f}(1-\gamma(\alpha))} {1- (N_c\alpha /2\pi )} \, ,
\label{NSVZ}
\end{eqnarray}
where $\gamma(\alpha)$ is  the anomalous dimension  of
the matter field,
\begin{equation}
\gamma(\alpha) =
 - \frac{N_c^2 - 1}{2 N_c}\frac{\alpha}{\pi}
 +  O(\alpha^2) \, .
\label{gam}
\end{equation}
The critical value of the coupling constant $\alpha_*$
is determined  by the zero of the $\beta$ function,
\begin{eqnarray}
\gamma (\alpha_* )= \gamma_{*} = 1 - 3\frac{N_c}{N_f}\, , \,\,\,
\alpha_* <
\frac{2\pi}{N_c}\, .
\label{alphastar}
\end{eqnarray}

According to Ref. \cite{S2} two dual theories, to be discussed below,
have the following content: the first one (``electric") has
$SU(N_c)$ gauge group and $N_f$ massless flavors while its dual  theory
(``magnetic") has $SU(N_f-N_c)$ gauge group, the same number $N_f$  of
massless flavors (but with different $U(1)$ quantum numbers) plus
$N_f^2$ colorless massless ``meson" fields \cite{S2}. The electric and
magnetic theories are supposed to have one and the same infrared limit
(although their behavior at short distances is distinct; see below).
Moreover, the  electric theory is weakly coupled near the right edge of
the window where  the magnetic one is strongly coupled, and {\em vice
versa}.

The main tool used in \cite{S2} for establishing the infrared
equivalence is the 't Hooft consistency condition \cite{thooft}. As was
first noted in \cite{dolgov} the chiral anomaly implies the existence
of the infrared singularities in the matrix elements of the axial
current (and the energy-momentum tensor) which are fixed unambiguously.
Therefore, even if we do not know how to calculate in the infrared
regime the  infrared limit of the theory should be arranged in such a
way as to match  these singularities.

The standard consideration is applicable only  to the so called
external anomalies. One  considers the currents (corresponding to global
symmetries) which are non-anomalous in the theory {\em per se},  but
acquire anomalies in  weak external backgrounds. For instance, in  QCD
with several flavors the singlet axial current is internally anomalous
-- its divergence is proportional to $G\tilde G$ where $G$ is the gluon
field strength tensor. The non-singlet currents are non-anomalous in
QCD but become anomalous if one includes the photon field, external with
respect to QCD. The anomaly in the singlet current does not lead to  the
statement of the infrared singularities in the current while the
anomaly in the non-singlet currents does. Thus, for the 't Hooft
matching one usually considers only the set of external anomalies.

At first sight in the conformal window the set of the external
anomalies includes extra currents due to the vanishing of the $\beta$
function  at the conformal point, $\beta(\alpha_*)=0$. In the framework
of supersymmetry it means that the trace of the stress tensor
$\theta^\mu_\mu$ vanishes as well as the divergence of some axial
current entering the  same supermultiplet as the stress tensor
$\theta_{\mu\nu}$. Therefore,  the idea that immediately comes to one's
mind is that the standard  matching conditions should be supplemented
by the new singlet axial current. Actually, this was the starting point
in the first version of this paper. The point is false, however.

Our analysis shows that:

\noindent (i) The number of the matching conditions at the conformal
point is  not expanded and is the same as in Ref. \cite{S2}.

\noindent (ii) However, what changes is the form of the conserved $R$
current,  to be used in the matching conditions; coefficients in the
definition of  the conserved $R$ current are $\alpha$ dependent, i.e.
are affected by  higher loops.

\noindent (iii) Although the $R$ current is different from the naive
one  (where
$\alpha$ is set equal to zero) {\it consequences} for the matching
conditions and superpotentials
  remain intact provided one takes into  account higher order
corrections consistently everywhere, together with the specific  form
of the NSVZ $\beta$ function. The crucial observation is the fact  that
the conserved $R$ current which includes all orders in coupling
constants still yields the 't Hooft consistency conditions with no
higher loop corrections. The fact that higher orders in $\alpha$ have
no impact in   some relations is due to  the existence of a new type of
holomorphy  in the effective Lagrangian for the anomalous triangles in
external fields.

The paper is organized as follows. In Sec. 2 we briefly review  those
results of Ref. \cite{S2} which are relevant for our analysis,
introducing notations to be used throughout the paper. In Sec. 3  we
discuss the construction of the conserved (anomaly free) $R$  current
to all orders in the coupling constants. Sec. 4 treats the anomaly
matching conditions at the multi-loop level. It is shown here that the
higher order corrections present in the $R$ current are  canceled in
the 't Hooft consistency conditions for the external backgrounds. Sec.
5 explains the cancellation  of the  higher order corrections in the
triangles for the $R$ current  to the baryon currents. In Sec. 6 we
discuss the selection rules for the  superpotentials. In Sec. 7 we
comment on inequivalence of the electric and magnetic theories at short
distances. Sec. 8 is devoted to incorporating the Yukawa  couplings in
the analysis of the infrared fixed points. The anomalous dimension of
the $M$ field is derived here from the requirement of  the conformal
symmetry in the infrared limit.

\section{One-loop anomaly matching condition}

The action  of the electric  theory is

\begin{eqnarray}
S =  \frac{1}{2g^2} \int d^4 x d^2 \theta ~ {\rm Tr}\,W^2
 +
\frac{Z}{4}
  \sum_{f}  \int d^4 x d^4 \theta ~
\left(\bar{Q}^{\dagger}_f e^{V}\bar{Q}_f
 + Q^{\dagger}_{f} e^{-V}Q_f \right)
\label{SUSYaction2}
\end{eqnarray}
where $Q_f$ and ${\bar Q}_f$  are the matter chiral
superfields in  the
$N_c$ and ${\bar N}_c$ color representations, respectively. The
subscript $f$ is the flavor index running from 1 to $N_f$.  The theory
has the following  global symmetries free from the internal  anomalies:
\begin{eqnarray} SU(N_f)_L\times SU(N_f)_R \times U(1)_B \times U(1)_R
\end{eqnarray} where the quantum numbers of the matter multiplets with
respect to these symmetries are as follows

\begin{center}
\begin{tabular}{|c|c|c|c|c|}\hline
 $ ~~$~~~  & ~~ $SU(N_f)_L$ ~~ & ~~$SU(N_f)_R$ ~~ &~~ $U(1)_B$ ~~ &
{}~~$U(1)_R$~~ \\\hline
$Q$ & $ N_f$  & 0 & 1 &  $ (N_f - N_c)/N_f$ \\
$\bar{Q}$ & 0 & $\bar{N}_f$ & -1 &$ (N_f -  N_c)/N_f$\\
\hline
\end{tabular}
\end{center}
\begin{center} Table 1
\end{center}

The $SU(N_f)_L\times SU(N_f)_R \times U(1)_B $ transformations   act
only on the matter fields in an obvious way, and do not affect the
superspace coordinate $\theta$. As for the extra global symmetry
$U(1)_R$ it is defined in such a way that it acts nontrivially on the
supercoordinate $\theta$ and, therefore, acts differently on the  spinor
and the scalar or vector components of superfields. The $R$ charges  in
the Table 1 are given for the lowest component of the chiral
superfields.

The notion of the
$R$ symmetries was originally introduced in Ref.~\cite{Salam}. The
$R$ current considered in Ref.~\cite{S2} is a conserved current that is
free from the triangle anomaly at the one-loop level.
At the classical level there are two conserved axial currents.
 One of them  -- sometimes the corresponding symmetry is called
$R_0$ --  is the axial current entering the same supermultiplet as the
energy-momentum tensor and the supercurrent. The $R_0$  current is
classically conserved if all matter fields are massless; at the quantum
level, generically, it acquires the internal anomaly proportional to
$\beta (\alpha) G\tilde G$, see  Refs. \cite{CPS,Clark,Shif1}.  Another
one is the flavor singlet current of the matter field.  The anomaly of
the latter current is purely one-loop. First it  was shown in
Ref.~\cite{CPS}, later an independent analysis of  this anomaly was
carried out by Konishi {\it et al}~\cite{Konishi,KONS}. The
corresponding expression is usually called the Konishi relation.

Seiberg's $R$ charge refers to a combination of these two
currents chosen  in such a way as to ensure cancellation of the
internal  anomaly at one loop.(Let us parenthetically note the
simplest example of conserved $R$ current in the Abelian gauge
theory was found long ago in Ref.~\cite{CPS}.)  This nonanomalous
$R$ symmetry   transforms  superfields in the following way:
$$
W(\theta) \rightarrow e^{-i\epsilon} W( e^{i\epsilon}\theta) \, ,
$$
\begin{eqnarray} Q(\theta)\rightarrow e^{i\epsilon(N_c-N_f)/N_f} Q(
e^{i\epsilon}\theta)\, ,
\;\;\;\;
\bar{Q}(\theta)\rightarrow e^{i\epsilon(N_c-N_f)/N_f}\bar{Q}(
e^{i\epsilon}\theta) \, .
\end{eqnarray}
In the component form  the $R$ transformations are
\begin{eqnarray}
\lambda\rightarrow e^{-i\epsilon} \lambda\,, ~~~
\psi(\bar{\psi})\rightarrow e^{i\epsilon N_c/N_f} \psi(\bar{\psi}),
{}~~~\phi(\bar{\phi})
\rightarrow e^{i\epsilon (N_c-N_f)/N_f}\phi(\bar{\phi})
\end{eqnarray}
where $\phi(\bar{\phi})$ and $\psi(\bar{\psi})$ are the
scalar and fermion components of chiral superfields $Q$ and
$\bar{Q}$.

  The conserved $R^S$ current is defined in \cite{S2} as
\begin{eqnarray}
R_{\alpha\dot{\alpha}}^{S} = \, \frac{2}{g^2} {\rm
Tr}\, (\lambda_{\dot{\alpha}}^\dagger\lambda_{\alpha}) -
\frac{N_c}{N_f} \sum_f \psi_{\dot{\alpha}}^{f\dagger}
\psi_{\alpha}^f -\frac{N_c}{N_f} \sum_f
\bar{\psi}_{\dot{\alpha}}^{f\dagger}
\bar{\psi}_{\alpha}^f
\label{Rcurrent}
\end{eqnarray}
where $\alpha$ and $\dot{\alpha}$ are the standard
spinor indices
and ${\rm Tr}\, (\lambda_{\dot{\alpha}}^\dagger\lambda_{\alpha})  =
 \frac{1}{2}\lambda_{\dot{\alpha}}^{\dagger\,a}
\lambda_{\alpha}^a$. Here and below the contribution of the scalars  in
the currents is consistently omitted.

It was suggested \cite{S2} that for $3N_c/2 < N_f < 3N_c$ there is
another (magnetic) theory with the  same  number of flavors $N_{f}$  but
different color group,  $SU(N_f-N_c)$, in which one has an additional
``meson" supermultiplets $M^i_j\;\,(i,j=1,...,N_f)$. (Below, to
distinguish the quark and gluon fields  of the magnetic theory from
those of the  electric one the former will be marked by tilde.)

The quantum numbers of the new chiral quark superfields and the meson
superfield $M$  with respect to the global symmetries
$ SU(N_f)\times SU(N_f) \times U(1)_B \times U(1)_R
$ are as follows:

\newpage

\begin{center}
\begin{tabular}{|c|c|c|c|c|}\hline
 $ ~~$~~~  & ~~ $SU(N_f)_L$ ~~ & ~~$SU(N_f)_R$ ~~ &~~ $U(1)_B$ ~~ &
{}~~$U(1)_R$~~ \\ \hline
$q $ & $ \bar{N}_f$  & $0$ & $N_c/(N_f-N_c)$ &  $ N_c/N_f$
 \\
$\bar{q}$ & $0$ & $N_f$ & $- N_c/(N_f-N_c)$ &$ N_c/N_f$
\\
$M$ &  $N_f$ & $\bar{N}_f$ & $0$ & $2 (N_f - N_c)/N_f$ \\
\hline
\end{tabular}
\end{center}
\begin{center} Table 2
\end{center} where the quantum numbers for meson field $M$ are defined
 from the superpotential
\begin{eqnarray} {\cal W} = f\,M^i_j q_i\bar{q}^j \, ,
\label{superpotentialM}
\end{eqnarray} and the conserved $R$ current must be  defined as

$$
{\tilde R}_{\alpha\dot{\alpha}}^{S} = \frac{2}{g^2} {\rm Tr}\,
(\tilde{\lambda}_{\dot{\alpha}}^\dagger\tilde{\lambda}_{\alpha}) -
\frac{N_f-N_c}{N_f} \sum_f \tilde{\psi}_{\dot{\alpha}}^{f\dagger}
 \tilde{\psi}_{\alpha}^f -\frac{N_f-N_c}{N_f} \sum_f
\tilde{\bar{\psi}}_{\dot{\alpha}}^{f\dagger}
\tilde{\bar{\psi}}_{\alpha}^f
$$
\begin{eqnarray}
+ \frac{N_f-2N_c}{N_f}{\rm
Tr}\,(\chi_{\dot{\alpha}}^\dagger\chi_{\alpha})
\label{dualRcurrent}
\end{eqnarray}
where $\tilde{\lambda}$ and $
\tilde{\psi}\;,{\tilde{\bar \psi}}$ are   dual gluino and  quarks and
$\chi$ are  the fermions from the supermultiplet
$M$.

Let us also note that if the number of the dual colors $N_f - N_c$ is
the same as $N_{c}$, i.e. $N_f = 2N_c$, then the $R$ charge of $M$ is
zero. At this point, $N_f = 2N_c$, the electric and magnetic   theories
look self-dual.  As we will see shortly, the actual situation is more
complicated. The fermions $\chi$ do not
 decouple from the $R$ current defined beyond one loop.

The electric and magnetic theories described above  are equivalent in
their respective  infrared (IR) conformal fixed points -- with   the
choice of the quantum numbers above  the highly non-trivial 't Hooft
anomaly matching  conditions for the  currents corresponding to
$ SU(N_f)\times SU(N_f) \times U(1)_B \times U(1)_R
$ are satisfied. If the $SU(N_c)$ theory is weakly coupled at the
conformal point $\alpha_*$  the dual $SU(N_f-N_c)$  theory  will be
strongly coupled, and it is only the anomaly relations  which can be
compared, because  they  can  be reliably calculated in both theories.
The presence of fermions from the  meson multiplet $M$ is   absolutely
crucial for this  matching. Specifically, one finds for the one-loop
anomalies in both theories \cite{S2}:
\begin{eqnarray}
SU(N_f)^{3}  &\rightarrow & N_{c}d^{(3)}(N_f)
\nonumber \\
SU(N_f)^{2}U(1)_R  & \rightarrow &   -\frac{N_c^2}{N_f}
d^{(2)}(N_f) \nonumber \\
SU(N_f)^{2}U(1)_B  &\rightarrow &  N_c
d^{(2)}(N_f) \nonumber \\
U(1)^2_B U(1)_R  &\rightarrow &  -2N_c^2
\nonumber \\ U(1)_R^3  & \rightarrow &  N_c^2 - 1 -
2\frac{N_c^4}{N_f^2} \nonumber \\
U(1)_R  & \rightarrow &  -N_c^2 - 1
\label{seibergmatching}
\end{eqnarray}
where the constants $d^{(3)}$  and $d^{(2)}$ were
introduced in
\cite{S2} and  are related to the traces of three and two  $SU(N_f)$
generators.

For example, in the $U(1)_R^3$ anomaly in the electric  theory the
gluino  contribution is proportional  to $N_c^2 - 1$ and  that of
quarks to $-(N_c/N_f)^3 2N_fN_c = - 2N_c^4/N_f^2$; altogether  $N_c^2 -
1- 2N_c^4/N_f^2$ as in (\ref{seibergmatching}). In the dual theory one
gets from gluino and quarks $\tilde{q},\,\, \bar{\tilde{q}}$  another
contribution, $(N_f-N_c)^2 - 1- 2(N_f-N_c)^4/N_f^2$. Then  the  fermions
$\chi$ from the meson multiplet $M$  (see Eq. (\ref{dualRcurrent}))
add extra $[(N_f-2N_c)/N_f]^3 N_f^2$, which is  precisely the
difference.

The last line in Eq. (\ref{seibergmatching}) corresponds to the anomaly
of the $R$ current in the background gravitational field,
\begin{eqnarray}
\partial^{\alpha\dot{\alpha}}J_{\alpha\dot{\alpha}}^{R}
\sim \epsilon_{\mu\nu\lambda\delta} R^{\mu\nu\sigma\rho}
R^{\lambda\delta}_{\sigma\rho} \, .
\end{eqnarray}
In the electric $SU(N_c)$ theory  the corresponding
coefficient is
$$ N_c^2-1 -
 2(N_c/N_f)N_c N_f = - N_c^2 - 1
$$
 while in the magnetic  theory
 it is $-(N_f-N_c)^2 - 1$ from quarks and gluinos and
 $[(N_f-2N_c)/N_f] N_f^2$ from the $M$ fermions, i.e.  in the sum
 again $- N_c^2 - 1$.

As was discussed above, in accordance with the standard logic, the  set
of the matching conditions above includes only those currents that do
not have internal anomalies. The number of the matching conditions  is
rather large and the fact they are satisfied with the given field
content is highly non-trivial.

\section{ Conserved $R$ currents}

In the consideration above it was crucial that there exists a singlet
axial current ($R$ current) whose conservation is preserved at the
quantum level. The particular form of the current (\ref{Rcurrent})
assumes that the coefficients are $\alpha$ independent numbers. We will
show below that this is not the case if higher loops are included and
we  will  determine the coefficients in terms of the anomalous
dimensions  $\gamma$ of the matter fields. This definition is
consistent with the fact that the anomaly in  the divergence of the
$R^0$ current is multi-loop.

Let us consider first the electric theory. At the classical level there
exist two conserved singlet currents.  The first one is the member of
the supermultiplet containing the stress tensor and the supercurrent
\cite{FZ}, it has the following  universal form:
\begin{eqnarray}
R^0_{\alpha\dot{\alpha}} =
\frac{2}{g^2} {\rm Tr}\,
(\lambda_{\dot{\alpha}}^\dagger\lambda_{\alpha})-
\frac{1}{3}\left( \sum_f \psi_{\dot{\alpha}}^{f\dagger}
\psi_{\alpha}^f
 +  \sum_f \bar{\psi}_{\dot{\alpha}}^{f\dagger}
\bar{\psi}_{\alpha}^f \right)\;.
\label{RT}
\end{eqnarray}
The current $R^0_{\alpha\dot{\alpha}}$ is the lowest
component of  the  superfield $J^0_{\alpha\dot{\alpha}}$ (see
Ref.~\cite{Clark}),
$$
 J^0_{\alpha\dot{\alpha}}=-\frac{2}{g^2}{\rm Tr}\,\left( W_{\alpha}
e^V  W^{\dagger}_{\dot{\alpha}} e^{-V} \right) +
\frac{Z}{12}\left\{\left[ (D_\alpha (e^{-V} Q)) e^V {\bar D}_{\dot
\alpha}(e^{-V} Q^{\dagger})
\right.\right.
$$
$$
\left. +  Q e^{-V} D_{\alpha}(e^V  {\bar D}_{\dot \alpha} (e^{-
V}Q^{\dagger}))
 +  Q {\bar D}_{\dot \alpha} (e^{-V} D_\alpha Q^{\dagger})  - (Q
\rightarrow Q^{\dagger}, V \rightarrow - V) \right]
$$
\begin{eqnarray}
\left. + (Q \rightarrow {\bar Q}\,,\; V \rightarrow -V) \right\}.
\label{super1}
\end{eqnarray}

 The current (\ref{RT}) corresponds to the transformation of the
superfields
$$
W(\theta) \rightarrow e^{3i\alpha} W( e^{-3i\alpha}\theta)\, ,
$$
\begin{eqnarray}
Q\left(\theta\right)\rightarrow e^{2i\alpha}
Q\left(e^{-3i\alpha}\theta\right)\, ,\,\,\,
\bar Q\left(\theta\right)\rightarrow e^{2i\alpha}
\bar Q\left(e^{-3i\alpha}\theta\right)\, .
\label{3-1}
\end{eqnarray}
In components this means
\begin{eqnarray}
\lambda\rightarrow e^{3i\alpha} \lambda, ~~~
\psi(\bar{\psi})\rightarrow e^{-i\alpha} \psi(\bar{\psi}),
{}~~~\phi(\bar{\phi})
\rightarrow e^{2i\alpha}\phi(\bar{\phi})\, ;
\end{eqnarray}
this symmetry exists in the presence of the
 Yukawa couplings of the form $c_{ijk}Q_i Q_j Q_k$.

The second classically conserved current,  $K_{\alpha \dot{\alpha}}$,
 which we will refer to as the Konishi current, is built from the
matter fields only:
\begin{equation}
K_{\alpha \dot{\alpha}} =
\frac{1}{3} \sum_f \psi_{\dot{\alpha}}^{f\dagger} \psi_{\alpha}^f
+\frac{1}{3} \sum_f \bar{\psi}_{\dot{\alpha}}^{f\dagger}
\bar{\psi}_{\alpha}^f\;.
\label{Kcurrent}
\end{equation}
Note, that although the Konishi current (\ref{Kcurrent})
superficially looks identical to the second term in equation (\ref{RT}),
actually  they are different -- the (omitted) contributions of the
scalars in
 (\ref{RT}) and (\ref{Kcurrent}) are different.   The current
(\ref{Kcurrent})  is the lowest component of the  superfield
$$
-\frac{Z}{12}\left\{\left[ (D_\alpha (e^{-V} Q)) e^V {\bar D}_{\dot
\alpha}(e^{-V} Q^{\dagger})  - \frac{1}{2}Q e^{-V} D_{\alpha} (e^V
{\bar D}_{\dot \alpha} (e^{-V}Q^{\dagger})) - \frac{1}{2} Q {\bar
D}_{\dot \alpha} (e^{-V} D_\alpha  Q^{\dagger})\right.
 \right.
$$
\begin{equation}
\left.\left. - (Q \rightarrow Q^{\dagger}, V \rightarrow - V) \right]
+ (Q \rightarrow {\bar Q}\,,\; V \rightarrow -V) \right\}.
\label{super5}
\end{equation}

The  Konishi current corresponds to the transformation
 of the superfields:
\begin{equation} W(\theta) \rightarrow  W(\theta),~~~
Q\left(\theta\right)\rightarrow e^{i\beta} Q\left(\theta\right), ~~~
\bar Q\left(\theta\right)\rightarrow e^{i\beta}
\bar Q\left(\theta\right)
\end{equation}

Both currents $R^{0}_{\mu}$ and $K_{\mu}$ have anomalies at the
quantum level. The anomaly in the $R^0$ current is multi-loop   and in
the $K$ current is one-loop. In the operator form   the anomalies can
be  written as follows \cite{Shif1}:
\begin{equation}
\bar{D}^{\dot{\alpha}} J^{0}_{\alpha\dot{\alpha}} =
 - \frac{1}{24} D_{\alpha}\left[
\frac{3N_c-N_f}{2\pi^2} {\rm Tr} \,W^2 +
\gamma Z \bar{D}^2 \sum_{f}
\left({Q}^\dagger_f e^{V}{Q}_f
 + \bar{Q}^\dagger_{f} e^{-V}\bar{Q}_f \right)\right]
\label{A1}
\end{equation}
and
\begin{equation}
 \bar{D}^2 Z \sum_{f}
\left({Q}^\dagger_f e^{V}{Q}_f
 + \bar{Q}^\dagger_{f} e^{-V}\bar{Q}_f \right) =
 \frac{N_f}{2\pi^2} {\rm Tr}\, W^2
\label{A2}
\end{equation} Here the anomalous dimension $\gamma$ is defined as
$$
\gamma = - \, \frac{d \,\ln Z}{d \,\ln \mu}\;,
$$
and in one loop is given by equation (\ref{gam}). Equation
(\ref{A2}) is the Konishi anomaly \cite{Konishi}-\cite{KONS}.  The
second term  in the right-hand side of equation (\ref{A1})   is due to
higher-loop  effects and represents the violation  of the holomorphy of
the effective Lagrangian.

 By virtue of the Konishi anomaly (\ref{A2}) the second term  in the
right-hand side of equation (\ref{A1}) is transformed into the same
gauge anomaly. The corresponding divergence in the $R^0$  current looks
as follows:
\begin{eqnarray}
\partial^{\mu}R^{0}_{\mu}  = \frac{1}{48\pi^2}
\,\left[3N_c -  N_{f}(1-\gamma)\right] \,
G^{a}_{\mu\nu}\tilde{G}^{a}_{\mu\nu}
\label{DR}
\end{eqnarray}
where the coefficient in the square bracket is the
numerator of the  NSVZ $\beta$-function (\ref{NSVZ}). The denominator
will  appear  after taking the matrix element of the operator
$G\tilde{G}$.

 Let us write down in parallel the anomaly in the matter current
\begin{eqnarray}
\partial^{\mu} K_{\mu}  = \frac{1}{48\pi^2}
\, N_{f} \,  G^{a}_{\mu\nu}\tilde{G}^{a}_{\mu\nu}
\label{DK}
\end{eqnarray}
{}From these anomalies we easily recover the form of the
{\it only}
 conserved $R$ current in the theory
\begin{equation} R_{\mu} = R_{\mu}^{0} + \left[1 -\frac{3N_c}{N_f}
-\gamma
\right] K_{\mu}\;.
\label{Rc}
\end{equation}
One  could give an alternative derivation of the very
same conserved
$R$ current considering the  mixing matrix we will between $R^0$ and
$K$ currents which arise already at the one-loop level. Diagonalizing
the  mixing  we can find two  renormalization group (RG) invariant
currents, one of which coincides  with the conserved $R_{\mu}$ and the
second one, which  is not
 conserved, is $(1-\gamma) K_{\mu}$ (see Ref. \cite{Shif1} for
details).

 Consider now  how the conserved current $R_{\mu}$ looks like in  two
limits. In the conformal point, when $\alpha = \alpha_{*}$ the
coefficient in front of $K_{\mu}$  vanishes (see Eq.~(\ref{alphastar})).
Thus in the infrared  limit the $R$ current flows to
$R^0$. On the  other hand in the  extreme ultraviolet ($UV$) limit
$\alpha(\mu)
\rightarrow 0$ (i.e.  $\gamma(\alpha) \rightarrow 0$)  the $R$  current
flows to  the  Seiberg $R^S$ current (\ref{Rcurrent}).  Therefore the
genuine
$R$ current interpolates between the $R^S$  and $R_{0}$ currents.

Keeping in mind that in the magnetic theory there are two distinct
matter fields with a superpotential, let us generalize the procedure of
construction of the conserved $R$ current  to the case with some
number of matter superfields $S_i$ and a nonvanishing superpotential
$\cal W $.

The definition of the $R^0$ current is general since it has a
geometrical nature,
\begin{eqnarray}
\tilde{R}^{0}_{\alpha\dot{\alpha}} = \frac{2}{g^2} {\rm Tr} \,
(\tilde{\lambda}_{\dot{\alpha}}^\dagger\tilde{\lambda}_{\alpha})  -
\frac{1}{3} \sum_i\psi_{\dot{\alpha}}^{i\dagger}
\psi_{\alpha}^i \;,
\label{R1}
\end{eqnarray}
where $\psi^i$ is the fermionic component of the chiral
superfield
$S^i$. In the presence of the superpotential $\cal W$ there are two
sources  of the current nonconservation: the first source is possible
classical nonconservation due to ${\cal W} \neq 0$, the second one is
the  quantum anomaly. In the superfield notations we have the following
generalization of the equation (\ref{A1}):
\begin{equation}
\bar{D}^{\dot{\alpha}} J^{0}_{\alpha\dot{\alpha}} =
 \frac{1}{3} D_{\alpha}\left\{\left[3{\cal W} - \sum_i S_i \,
\frac{\partial {\cal W}}{\partial S_i}\right]
 - \left[\frac{b}{16\pi^2} {\rm Tr}\, W^2 + \frac{1}{8} \sum_i
\gamma_i Z_i \bar{D}^2
\left({S_i}^\dagger e^{V}S_i\right) \right] \right\}
\label{J1}
\end{equation}
where $b=3N_c-\sum_i T_i$ is the first coefficient of
the $\beta$  function and invariants $T_i$ characterize the gauge group
representation of  the field $S_i$ (they are defined as ${\rm Tr}\,t^a
t^b= T_i \delta^{ab}$  where
$t^a$ are the matrices of the group generators).

We can also  construct the Konishi current
\begin{equation}
K^i_{\alpha \dot \alpha} = \frac{1}{3}
\psi_{\dot{\alpha}}^{i\dagger}\psi_{\alpha}^i\;.
\label{K1}
\end{equation}
for each superfield $S_i$. The divergence of this
current is given by the generalized Konishi  relation:
\begin{equation}
\frac{1}{8} \bar{D}^2
\left( Z_i {S_i}^\dagger e^{V}{S}_i \right) =
\frac{1}{2}S_i \, \frac{\partial {\cal W}}{\partial S_i} +
\frac{T_i}{16\pi^2} {\rm Tr}\, W^2 \;.
\label{K2}
\end{equation}
(A comment on the literature: the anomaly in the current
(\ref{J1})  was expressed in terms of anomalous dimensions $\gamma_i$ in
Ref.~\cite{Clark,Shif1}; for a recent instructive discussion which
includes the classical nonconservation
\cite{FZ}, see Ref. \cite{LS}. )

Let us look for the conserved $R$ current as a linear combination of
the  currents $R^0_{\alpha \dot \alpha}$ and
$K^i_{\alpha \dot\alpha}$~:
\begin{equation}
R_{\alpha \dot \alpha}=R^0_{\alpha \dot \alpha}+c_i
K^i_{\alpha \dot \alpha}\;.
\label{RM}
\end{equation}
The divergence of this current can be immediately found
from Eqs. (\ref{J1}) and (\ref{K2}),
\begin{equation}
\partial_\mu R^\mu = -\frac{4}{3}\left\{\left. \left[ 3{\cal W} -
\sum_i S_i \,
\frac{\partial {\cal W}}{\partial S_i} \left( 1+\frac{c_i +
\gamma_i}{2}\right)\right]\right|_G - \frac{{\rm Tr}\, W^2|_G}
{16\pi^2} \left[ b +\sum_i T_i\,(c_i +
\gamma_i)\right]\right\}
\;.
\label{Rdiv}
\end{equation}
where the subscript $G$ marks the $G$ component of the
chiral  superfield, in particular, ${\rm Tr}\,
W^2|_G=G^a_{\mu\nu}{\tilde G}^a_{\mu\nu}/4$. All terms proportional to
$\gamma_i$ occur by virtue of the  substitution of the Konishi relation
(\ref{K2}) into (\ref{J1}).

For the conserved $R$ current to exist both terms in Eq. (\ref{Rdiv}),
the superpotential term and the one with $W^2$, must vanish. All  higher
order corrections reside in the anomalous dimensions $\gamma_i$  which
depend on the gauge coupling constant $\alpha$ and the constants in the
superpotential ${\cal W}$.

Let us first omit these higher order corrections, i.e. put
$\gamma_i=0$. For a generic superpotential there may be no conserved
currents at  all. This situation is of no interest to us, so we assume
that one (or more) conserved currents exist. Let us denote by
$c_i^{(0)}$ the set of the coefficients
$c_i$ ensuring the vanishing of $\partial R$ at $\gamma_i=0$. (These
coefficients $c^{(0)}_i$ are rational numbers which in many cases  were
found by Seiberg {\it et al.})

Now let us switch on the higher order corrections, $\gamma_i \neq  0$.
It is crucial that the anomalous dimensions $\gamma_i$ appear only in
the combination $c_i + \gamma_i$. This means that the coefficients
$c_i$ ensuring the current conservation at $\gamma_i\neq 0$ are
different from
$c^{(0)}_i$ only by a shift by $(-\gamma_i)$,
\begin{equation}
c_i=c^{(0)}_i - \gamma_i
\label{c_i}
\end{equation}

Equation (\ref{Rc}) given above is a particular example of this  general
result with $c^{(0)}=1-(3N_c/N_f)$. As another illustration let us
consider the magnetic theory. In this case we have two Konishi currents,
\begin{equation}
\tilde{K}_{\alpha \dot{\alpha}}^q =
\frac{1}{3} \sum_f \tilde{\psi}_{\dot{\alpha}}^{f\dagger}
\tilde{\psi}_{\alpha}^f +\frac{1}{3} \sum_f
\tilde{\bar{\psi}}_{\dot{\alpha}}^{f\dagger}
\tilde{\bar{\psi}}_{\alpha}^f\;\;,
\; \;\;\;\;\;
\tilde{K}_{\alpha \dot{\alpha}}^M =
\frac{1}{3} {\rm Tr}\, (\chi^{\dagger}_{\dot{\alpha}}\chi_\alpha)
\label{Ktilde}
\end{equation}
where the notations have been introduced in Sec. 2.
Equation (\ref{dualRcurrent}) gives the values of $c_i^{(0)}$ in this
case:
\begin{equation}
c_q^{(0)}=-\frac{1}{2}c_M^{(0)}=\frac{3N_c-2N_f}{N_f}\;.
\label{cqcm}
\end{equation}
Higher order corrections will change these coefficients to
\begin{equation}
c_q=\frac{3N_c-2N_f}{N_f} - \gamma_q\;\;;\;\;
c_M=-2\frac{3N_c-2N_f}{N_f}- \gamma_M\;.
\label{cqcmcor}
\end{equation}
where $\gamma_q$ and $\gamma_M$ are the anomalous
dimensions  of the fields
$q$ and $M$ respectively. Thus, the extra terms in the $\tilde R$
current  as compared to $R^S$ one (see Eq.~(\ref{dualRcurrent})) are
\begin{equation}
\tilde R_{\alpha \dot{\alpha}} - R^S_{\alpha \dot{\alpha}} = -\gamma_q
\tilde{K}_{\alpha\dot{\alpha}}^q  - \gamma_M
\tilde{K}_{\alpha\dot{\alpha}}^M \;.
\label{extra}
\end{equation}

 One more comment concluding this section.  The conserved $R$  current
is in the same  supermultiplet with the stress-energy tensor
$\theta_{\mu\nu}$   {\it only} in the  infrared limit. Thus the
relation $D = (3/2)R$   between the dimension $D$ and the chiral  $R$
charge of the  chiral superfield  is valid only for the  IR fixed
point,  but not for the  UV one.

\section{Cancellation of higher-loop corrections for the external
anomalies}

Now, when the conserved $R$ current is constructed we can proceed  to
discussing the anomalies of this current in the background of weak
external fields. As was mentioned in Sec. 1, these anomalies, via the
't Hooft consistency conditions, constraint the infrared behavior of
the  theory, and, thus, crucial in establishing the electric-magnetic
duality. In Ref.
\cite{S2} the anomaly relations were analyzed at the one-loop level.
Since the duality can take place only in the conformal points where the
coupling constants are not small a consideration of higher-loop
corrections is absolutely crucial. As we see, for instance, from Eq.
(\ref{J1}), generally speaking, higher-loop corrections are present.
Below we  will demonstrate that in the external anomalies for the $R$
current  constructed above all higher order contributions cancel out.

To warm up we begin our consideration of the multi-loop effects
starting from the example of
$U(1)_{R}U(1)_{B}^2$ triangle. The definition of the baryon  charges
for the electric theory is given in section 2. We have $\alpha$
corrections both in the definition of the  current $R$ (see Eq.
(\ref{Rc})) and in the anomalous triangle.

 If we introduce an external field $B_{\mu}$ coupled to the  baryon
current then the anomaly for the $R^0$ current in the electric theory
takes the form similar to equation (\ref{DR}),
\begin{equation}
\partial^{\mu} R^0_{\mu} = - \frac{1}{24\pi^2} N_f\,N_c\,(1-\gamma)\,
B_{\mu\nu}\tilde{B}_{\mu\nu}
\label{RBB}
\end{equation}
 where $B_{\mu\nu} = \partial_{\mu}B_{\nu} -
\partial_{\nu}B_{\mu}$. Let us emphasize that $\alpha$
 corrections to the anomalous triangle do not vanish and
 enter through the anomalous  dimension $\gamma(\alpha)$ which is, in
turn, related to the second term in the square brackets in  Eq.
(\ref{A1}).
 As far as the matter current $K_{\mu}$ is concerned
 here $\alpha$ corrections to the bare triangle vanish,
\begin{equation}
\partial^{\mu} K_{\mu} = \frac{1}{24\pi^2} N_f\,N_c\,
B_{\mu\nu}\tilde{B}_{\mu\nu}
\label{KBB}
\end{equation}
Assembling these two pieces together we find
\begin{equation}
\partial^{\mu} R_{\mu} = \frac{1}{16\pi^2} (- 2N_{c}^2)\,
B_{\mu\nu}\tilde{B}_{\mu\nu} \;.
\label{fullRBB}
\end{equation}
One can see that the external anomaly in the $R$ current has no
$\alpha$ corrections -- those coming from the definition of the  current
are canceled by the corrections in the triangle   containing
$R^0_{\mu}$. Effectively the naive one-loop   anomaly is preserved  in
the full multi-loop calculation in the case at hand.

As a matter of fact it is not difficult to generalize the assertion
above to the case of arbitrary number of matter fields with a
superpotential (e.g. in the magnetic theory we have two distinct sets
of the matter  fields  -- the magnetic quarks in  the fundamental
representation of the gauge group and color  singlets).

Since the background field $B_\mu$ can be treated as an additional
gauge  field what is to be done, Eq.~(\ref{Rdiv}) must be  supplemented
by the extra term  proportional to $W^2_B|_G = B_{\mu \nu} {\tilde
B}^{\mu \nu}$. Namely,
\begin{equation}
\partial^{\mu} R_{\mu} = \frac{B_{\mu \nu} {\tilde B}^{\mu
\nu}}{32\pi^2}
\sum_i T^B_i (-1 + c_i + \gamma_i)
\label{deltaR}
\end{equation}
where $T^B_i$ is equal to the trace of the square of the
generators of  the baryon charge, i.e. $T^B_i= B_i^2$ ($B_i$ is the
baryon charge of the  field
$S_i$). The analogy with the second term in Eq. (\ref{Rdiv}) is quite
transparent. The color generators are substituted by those of the
baryon charge, and the coefficient $b$ in the square brackets is
substituted  by $(-\sum T_i^B)$.

The coefficients $c_i$ in Eq.~(\ref{Rdiv}) are chosen in such a way
that  the right-hand side of Eq.~(\ref{Rdiv}) vanishes. This vanishing
is  achieved provided
$c_i + \gamma_i=c_i^{(0)}$, see Eq. (\ref{c_i}). It is important that
the very same combination, $c_i + \gamma_i$, enters Eq.~(\ref{deltaR}).
This means that the external baryonic anomaly reduces to
\begin{equation}
\partial^{\mu} R_{\mu} = \frac{B_{\mu \nu} {\tilde B}^{\mu
\nu}}{32\pi^2}
\sum_i T^B_i (-1 +c_i^{(0)})
\label{deltaR1}
\end{equation}
Thus, the higher order corrections obviously drop out.

If the gravitational field is considered as external the calculation of
the corresponding triangle is very similar to that in the external
$B_\mu$ field. The only difference compared to
the case of the baryon current is the substitution of $T^B_i$ by
unity,
\begin{equation}
\partial^\mu (\sqrt{-g}R_{\mu})\,=\,\frac{1}{192\pi^2}\left[(N_c^2-1)  -
\frac{1}{3}\sum_i (-1 +c_i^{(0)}) \right]
\epsilon_{\mu\nu\lambda\delta} R^{\mu\nu\sigma\rho}
R^{\lambda\delta}_{\sigma\rho} \; ,
\label{Rgrav}
\end{equation}
In the electric theory $c_i^{(0)}=1 - (3N_c/N_f)$ while
in the  magnetic theory $N_c$ is substituted by $N_f - N_c$ and
$c_{q,M}^{(0)}$ are  given in Eq.  (\ref{cqcm}).

The arguments can be of course repeated for the $U(1)_R\,  SU(N_f)^2$
anomaly.  What remains to be discussed is the anomalous
$U(1)_R^{3}$ triangle. This case is harder to consider along the lines
presented here. However, in the next section we will give some
arguments of a more general nature indicating that this anomaly is also
one-loop.

\section{Cancellation of higher orders and holomorphy in the  external
field}

Now we will show that the result derived above -- cancellation of the
higher order corrections in the external anomalies -- has a very
transparent interpretation in the language of the effective action in
the external field. As an example of the external field one can keep in
mind  the field $B_\mu$ interacting with the baryonic charge,
gravitational  field  and so on. The Wilson effective action is as
follows:
$$
S_W(\mu) =  \frac{1}{16\pi^2}
\left[\frac{8\pi^2}{g_0^2} -\left( 3T_G \ln \frac{M_0}{\mu} - \sum_i
T_i \ln\frac{M_i}{\mu}
\right)\right]
\int d^4 x d^2
\theta ~ {\rm Tr}\, W^2 +
$$
$$
\frac{1}{16\pi^2} \left[ \sum_i T_i^{(ext)} \ln \frac{M_i}{\mu} \right]
\int d^4 x d^2 \theta ~  W^2_{ext} +
$$
\begin{equation}
\sum_i \frac{Z_i}{4}
 \int d^4 x d^4\theta ~{S_i}^{\dagger} e^{V} {S_i}+
\left\{ \int d^4 x d^2\theta ~ {\cal W} (S) + h.c.\right\}
\label{SW1}
\end{equation}
where $1/g_0^2$ is the inverse bare coupling constant
and $\mu$ is  the normalization point, the coefficients $T_i$ are
defined through the generators of the gauge group in the representation
$i$, ${\rm Tr} (t^a t^b) =T_i \delta^{ab}$, ($T_G$ is the invariant
$T_i$ in the adjoint representation). The masses $M_0$ and $M_i$ are
the  regulator  masses (one can keep in mind the supersymmetric
Pauli-Villars regularization). In the covariant computation $M_0$ is
the mass of  the (chiral) ghost regulators, $M_i$ is the mass of the
$S_i$ field  regulator. Usually, it is assumed that  all regulator
masses are the same,
$M_0 = M_i$. We keep them different for the purposes which will  become
clear shortly. Finally, $W_{ext}$ is a superfield generalizing the
stress tensor of the external gauge field in the same way as $W$
generalizes
$G_{\mu\nu}$ (in the previous section where the external baryonic
current was considered as an example we dealt with $W_B$), the
coefficients $T_i^{ext}$ are defined similar to
$T_i$ for the generators corresponding to the interaction with the
external field.  The superpotential ${\cal W} (S_i)$ may or may not be
present in  each particular model.

 The property of holomorphy in the
 Wilson effective action means that the coefficients in front of
 $W^2$ and $W_{ext}^2$ are given by one loop; higher order  corrections
 in the coupling constants are absent. Higher orders enter only the
$Z$ factors; in taking the background field matrix elements of the last
term in  Eq. (\ref{SW1}) higher
 orders in $Z$ penetrate the answer.

Taking matrix elements of the operator action $S_W$ we proceed to  the
c-number functional $\Gamma$, the generator of 1PI vertices. Let us
first discuss what happens at one-loop level. Then the last two terms in
Eq.~(\ref{SW1}) are irrelevant for the issue of anomalies under
discussion, and the one-loop description is given by
$$
\Gamma^{one-loop} = \frac{1}{16\pi^2}
\left[\frac{8\pi^2}{g_0^2} -\left( 3T_G \ln \frac{M_0}{\mu} - \sum_i
T_i \ln\frac{M_i}{\mu}
\right)\right]  \int d^4 x d^2 \theta ~ {\rm Tr}\, W^2 +
$$
\begin{equation}
\frac{1}{16\pi^2} \left[ \sum_i T_i^{(ext)} \ln \frac{M_i}{\mu} \right]
\int d^4 x d^2
\theta ~  W^2_{ext} \;.
\label{gamma}
\end{equation}

{}From this expression one can easily read off anomalies by varying  the
regulator masses. For instance, the anomaly of $R_0$ current is
obtained by applying the operator
\begin{equation}
 M_0\frac{\partial}{\partial M_0} + \sum_i
M_i\frac{\partial}{\partial  M_i}
\label{dm}
\end{equation}
to the right-hand side of Eq.~(\ref{gamma}). The anomaly
in the  Konishi current $K_i$ (see Eq.~(\ref{K1})) is generated by
$M_i(\partial/\partial M_i)\Gamma^{one-loop}$.

Moreover, the same expression (\ref{gamma}) demonstrates the existence
of the  conserved current $R$. Indeed, the first term in
$\Gamma^{one-loop}$ is  invariant under the action of the operator
\begin{equation}
M_0\frac{\partial}{\partial M_0} + \sum_i
M_i\frac{\partial}{\partial  M_i} (1+c_i^{(0)})
\label{dm1}
\end{equation}
where coefficients $c_i^{(0)}$ were defined and
discussed in Sect. 3.  The non-invariance of the second term in
$\Gamma^{one- loop}$ under the action of the operator (\ref{dm1}) gives
the external anomaly of  the $R$ current.

Now what happens if we proceed to higher loops? The occurrence of  the
$Z$ factors in $S_W$ manifests itself in $\Gamma$ in the following way
\cite{Shif1}:
$$
\Gamma^{multi-loop} = \frac{1}{16\pi^2}
\left[\frac{8\pi^2}{g_0^2} -\left( 3T_G \ln
\frac{M_0}{(g_0/g)^{\frac{2}{3}}\,\mu} - \sum_i T_i
\ln\frac{M_i}{Z_i\,\mu}
\right)\right]
\int d^4 x d^2
\theta ~ {\rm Tr}\, W^2
 +
$$
\begin{equation}
\frac{1}{16\pi^2} \left[ \sum_I T_i^{(ext)} \ln \frac{M_i}{Z_i \mu}
\right]
\int d^4 x d^2
\theta ~  W^2_{ext} \;.
\label{gammamul}
\end{equation} We write down here only the part of $\Gamma$ containing
$W^2$  and
$W^2_{ext}$; the part with classical superpotential is omitted.  The
inclusion of higher orders resulted in substituting the regulator  mass
$M_i$ by $M_i/Z_i$. The role of $Z_i$ for the ghost regulator mass
$M_0$ is  played by
$(g_0/g)^{\frac{2}{3}}$ (see Ref.~\cite{Shif1}).

Invariance of the $W^2$ part of $\Gamma$ still persists. However,  the
transformation under which it is invariant corresponds now to the
action of the operator
\begin{equation}
{\cal M}_0\frac{\partial}{\partial {\cal M}_0} +
\sum_i  {\cal M}_i\frac{\partial}{\partial  {\cal M}_i} (1+c_i^{(0)})
\label{dm2}
\end{equation}
where
\begin{equation} {\cal M}_i=\frac{M_i}{Z_i}\;,\;\;\;\; {\cal
M}_0=\frac{M_0}{(g_0/g)^{\frac{2}{3}}}\;.
\label{mi}
\end{equation}
The application of operator (\ref{mi}) to the
$W^2_{ext}$ part of
$\Gamma$ yields the external anomaly. Since in this part $M_i$ is also
replaced  by
${\cal M}_i=M_i/Z_i$ it is clear that the external anomaly remains
one-loop.

A direct correspondence between the discussion of the external
anomalies in Sect. 4 and the one given in this section is quite clear.
However, the  arguments of this section help  demonstrate the general
nature of the  phenomenon. In particular it seems possible to deduce
that the anomaly of the type
$U(1)^3_R$ is also one-loop.

\section{Superpotentials}

We have demonstrated that the conserved $R_\mu$ current contains higher
order terms. The question arises about the selection rules
 for different terms in superpotentials which were obtained without
 these  complications \cite{ADS}.

The change of the form of the $R$ current does not mean that these
selection rules were incorrect. The same phenomenon of the cancellation
of  the higher order corrections as was described above  takes  place
in the transformation laws of the  chiral fields.

To elucidate this assertion it is instructive to analyze the form
factor  diagrams where the chiral matter scatters off the current $R$.
Taken at the vanishing momentum transfer these graphs yield the
$R$ charge of the matter field. At a naive level one would start from
the naive current $R\mid_{\gamma =0}$, with all $\gamma$ terms
discarded, draw the tree graph plus two  one-loop graphs (the diagram
with the vertex correction and the  diagram with the correction to the
external line), and then one would  conclude that the two one-loop
graphs cancel each other in the same  way similar diagrams cancel each
other in the electric current in QED. This naive conclusion would be
wrong! If the calculation is done  supersymmetrically, in terms of the
supergraphs, getting a  non-vanishing result for the sum of the two
one-loop graphs  mentioned above is inevitable. The residual sum of
these graphs is canceled, however, when one adds the $\gamma$ term of
the $R$ current as the vertex insertion in the tree graph. Effectively
this means that the $R$ charge of the chiral matter is determined by the
tree graph with $R\mid_{\gamma =0}$ at the vertex, i.e. coincides with
Seiberg's answer. In other words the  very same assertion can be
phrased as follows: the commutator of the $R$ charge with the
(bare) matter  fields
$S_i$ contains an anomalous part that cancels the $\gamma$ terms  in
the definition of the conserved $R$ charge. In particular, the
commutator with the modular fields
$$
\left[ R,\; Q^i_f\bar{Q}_{i\,f'}\right]
$$
remains the same as in the naive analysis with all $\gamma$'s set
equal to zero.

A nice illustration of how this cancellation works is provided by the
simplest supersymmetric model, the massless Wess-Zumino model. This
example was thoroughly analyzed in Ref.~\cite{Clark} and we summarize
here the basic points. The action of the model is
\begin{equation}
S_{WZ} = \frac{Z}{4}
 \int d^4 x d^4\theta ~    {S_0}^{\dagger} {S_0} +
\left\{\int d^4 x d^2\theta ~ f_0S_0^3 + h.c.\right\}
\, ,
\label{WZ}
\end{equation}
where $S_0$ is the bare field and $f_0$ is the bare
coupling constant. The $R^0$ current of this model is merely
$$
-\frac{1}{3} \psi_{\dot{\alpha}}
\psi_{\alpha}\, ,
$$
i.e. the lowest component of  the  superfield
$$
J^0_{\alpha\dot{\alpha}}=
\frac{Z}{12}\left\{\left[ (D_\alpha S)  {\bar D}_{\dot
\alpha}(S^{\dagger})  +  S  D_{\alpha}{\bar D}_{\dot \alpha} S^{\dagger}
 +  S {\bar D}_{\dot \alpha} D_\alpha S^{\dagger}  - (S\rightarrow
S^{\dagger}) \right]
 \right\}\, .
$$
The $R^0$ current is classically conserved. The corresponding field
transformation
$$
S(\theta ) \rightarrow e^{2i\alpha} S(e^{-3i\alpha}\theta )
$$
leaves invariant both the kinetic and the superpotential terms in the
Wess-Zumino action. However, since $R^0$ enters the same
supermultiplet as the energy-momentum tensor it conservation is  ruined
at the quantum level when loops are included. The non-conservation  is
associated  only with the occurrence of the $Z$ factor in the kinetic
term and can be interpreted as the anomalous non-invariance of the
kinetic term under the phase rotation generated by the $R^0$  current.
The superpotential term, written through the bare fields remains
invariant -- this follows from the famous non-renormalization theorem
\cite{GRS} which tells us that there are no  quantum corrections to the
superpotential term and, hence, no dependence of this term on the
normalization point $\mu$.
 The anomalous divergence has the form
$$
\partial_\mu R^0_\mu =\gamma \partial_\mu K_\mu
$$
where $K_\mu$ is the Konishi current of the Wess-Zumino model,
$$
K_\mu = \frac{1}{3} \psi_{\dot{\alpha}}
\psi_{\alpha}\, ,
$$
or the lowest component of the superfield
$$
-\frac{Z}{12}\left\{\left[ (D_\alpha S)  {\bar D}_{\dot
\alpha}S^{\dagger} - \frac{1}{2}S  D_{\alpha} {\bar D}_{\dot \alpha}
S^{\dagger} - \frac{1}{2} S {\bar D}_{\dot \alpha} D_\alpha
S^{\dagger})- (S \rightarrow S^{\dagger}) \right]
\right\} .
$$
Up to a sign, the fermion parts of both currents look identical.
Transferring $\gamma \partial_\mu K_\mu$ to the left-hand side of the
anomaly relation we get the conserved $R$ current. It is clear that the
role of the term $-\gamma K_\mu$ in $R_\mu$ is to kill the anomalous
non-invariance of the kinetic  term under the phase rotation. It has no
effect whatsoever on the transformation properties of the
superpotential term, which is absolutely clear from the derivation of
the anomaly. If we  forget altogether about this anomalous
non-invariance of the kinetic term (as one would do naively) and use
the naive charge (in this case this is just the $R^0$ charge) we arrive
at the correct  conclusion concerning the invariance of the
superpotential term.  This situation is quite general.

The important lesson stemming from  the analysis is as follows: the
selection rules obtained for the  superpotential terms from the naive
(one-loop) $R^S$ current are valid only provided we work in terms of
the bare  (unrenormalized) fields, so that all $Z$ factors reside in
the kinetic  terms.

\section{Comments on short  distance behavior in the electric and
magnetic theories}

In this section we comment further on  the question what is the precise
meaning of the duality between the electric and magnetic theories.
According to Ref. \cite{S2} it should  be understood as equivalence of
the infrared limits of both theories -- i.e. large distance  scaling of
all correlation functions with the same anomalous dimensions for
equivalent operators. Sometimes a  stronger conjecture is made in the
literature -- the full equivalence of the  electric and magnetic
theories, at all distances.
 Here we will consider corresponding  correlation  functions at short
distances in the  electric and magnetic theories and explicitly
demonstrate that they are different.

First of all, we must choose operators that correspond to each other  in
the electric and magnetic theories. No general relation between
respective operators is known. The correspondence between  operators in
both theories may  be complicated, even nonlocal. We can consider,
however, the Noether currents which in both theories correspond to  one
and the same symmetry. The simplest choice is the baryon number
current.

Due to asymptotic freedom we can compare the corresponding   correlation
functions at short distances. The leading asymptotics is that of the
corresponding free  theories, and the disbalance is obvious.  The only
subtle point which deserves discussion is the presence of  the
additional Yukawa  interaction  in the magnetic theory.

The standard  argument is as follows.  The Yukawa coupling  is not
asymptotically free; therefore,  at short distances it explodes, and
this explosion precludes one from calculating the short-distance
behavior  in the magnetic theory.

 In the next section we will  explicitly demonstrate that the Yukawa
coupling {\em is} asymptotically free in the conformal window, due  to
the contribution to the coupling renormalization coming from the
exchange of the gauge fields, at least in some domain of the  parameter
space. Moreover, the magnetic theory has a  nontrivial infrared fixed
point for {\em both} couplings -- the gauge and the Yukawa.  (Otherwise,
the corresponding theory would not be conformal at all, no scaling
would be achieved at large distances, and no conformal window would
exist.)

 Thus we will compare two asymptotically free theories at short
distances.  In the asymptotic regime when all couplings become
arbitrarily small  a straightforward calculation yields the two-point
correlation functions  for the  baryon currents at short distances,
\begin{equation}
\langle 0|\{ J_{B}^\mu (x) J_{B}^\mu (0)\}|0\rangle  =   C
\frac{2}{\pi^2} ~\frac{1}{x^6} + ...,~~~~~~
 x \rightarrow 0
\end{equation}
 where the coefficient $C$ depends on the theory.   In the electric
theory the baryon  charge  is $1$ for $Q$ and $-1$ for  $\bar{Q}$, so
the constant $C$ can be easily shown to  be
$$
 C_{E} = N_f N_c\, .
$$
In the magnetic theory the quark baryon charge is
 $N_c/(N_f - N_c)$ and  the constant  $C_{M}$ now is equal to
$$
C_{M} = \left(\frac{N_c}{N_f - N_c}\right)^2  N_f  (N_f - N_c) =
 \frac{N_c}{N_f - N_c} C_{E} \, .
$$
 Only for $N_f = 2N_c$ these constants are the same (which is evident
because in this case  the quark charges in the electric theory  are the
same as in  the magnetic one). However, even at $N_f = 2N_c$, when the
baryon two-point functions match, it is not difficult to show that the
short distance behavior of some other two-point  functions (e.g.
induced by the $SU(N_f)$  currents) does not match. Even if we are not
in the domain of the parameter  space where the coupling $f$ is
asymptotically free, still $f$ approaches  small values at intermediate
scales (if the gauge coupling is asymptotically  free which is
necessary for the consistency of the whole approach). This  means that
the disbalance between the corresponding correlation functions  in the
electric and magnetic theories can be established at this intermediate
scale.

Thus one can see that this is the general case -- the  short distance
behavior  of correlation functions is different in both theories.

\section{Conformal fixed points and the Yukawa couplings}

 Let us discuss now the impact of the  inclusion  of the Yukawa
interaction  ${\cal W} = f M^i_j q_i\bar{q}^j$ in the  action  of the
dual theory. For simplicity of notations we will write  that the gauge
group is some $SU(N_c)$ and will omit tilde in all fields.  The  problem
which we are interested in is the aspects of the behavior near the
infrared  fixed point.

The  action takes the form
\begin{eqnarray}
S =\frac{ Z_{q}(\mu)}{4}
  \sum_{f}  \int d^4 x d^4 \theta
\left(\bar{q}^{\dagger}_f e^{V}\bar{q}_f
 + q^{\dagger}_{f} e^{-V}q_f \right) +
\frac{Z_{M}(\mu)}{4}
   \int d^4 x d^4 \theta  ~ M^{\dagger} M + \nonumber \\
 \frac{1}{2g^2(\mu)} \int d^4 x d^2 {\rm Tr}\,\theta ~ W^2
 + \left[ f_0 \int d^4 x d^2 \theta ~M^i_j q_i\bar{q}^j
 + h.c. \right] ~~~~~~~~~~~~~~~~
\label{SUSYaction3}
\end{eqnarray}
where $Z_q$ and $Z_M$ are the quark and $M$ field $Z$
factors and
$f$ is the Yukawa coupling. The Yukawa interaction is the $F$ term and
due to the famous nonrenormalization theorem \cite{GRS} $f_0$   does
not depend on $\mu$; the  renormalized coupling is  unambiguously
defined as
\begin{eqnarray}
 f(\mu) = f_0/ (Z_{q}(\mu)\sqrt{ Z_{M}(\mu)})\, .
\end{eqnarray}
The renormalization group  equation for $f(\mu)$ then
has the form
\begin{eqnarray}
\frac{d f^2}{d \ln\mu} =  f^2 ~ \left[\gamma_{M}(\alpha,f)
 + 2 \gamma_{q}(\alpha,f)\right]
\end{eqnarray}
where the quark and meson anomalous dimensions are
defined  in  the  standard way
\begin{eqnarray}
\gamma_{q}(\alpha, f) = - d \ln Z_q/d \ln \mu,~~~~
\gamma_{M}(\alpha, f) = - d \ln Z_M/d \ln \mu \,  .
\end{eqnarray}

 If we believe in the  conformal window we must insist that  there is
an IR  fixed point for  both  couplings --   gauge and Yukawa.  Then
one of the conditions  of the IR fixed point  $(\alpha_{*}, f_{*})$
will be
\begin{equation}
\gamma_{M}(\alpha_{*}, f_{*})
 + 2 \gamma_{q}(\alpha_{*}, f_{*}) = 0
\label{curve1}
\end{equation}
(Let us parenthetically note that one can obtain the
condition  for the  zero of $\beta$ function (\ref{curve1})  by
studying the  multiplet
 of anomalies. This was done in a recent paper by  Leigh and  Strassler
\cite{LS} who discussed marginal  operators in different $N=1$ and
$N=2$ theories.)

 One can make  very general statements  about the asymptotic  behavior
of the $Z$ factors  in the infrared region.   The first one   is  that
$Z_q \rightarrow 0$  when approaching  the conformal point.  To see
this is indeed the case let us start from a particular example, $f =
0$.  Then
 $Z_{q}(\mu) =  Z_{q}(\alpha(\mu))$ because of the renormalizability  of
the  theory, and
\begin{eqnarray}
-\gamma_{q}(\alpha) =  \frac{d \ln
Z_{q}(\alpha(\mu))}{d \ln \mu} =
  \frac{ d \ln Z_{q}(\alpha)}{d  \alpha }
\frac{d \alpha(\mu)}{d \ln \mu} = \beta(\alpha)
\frac{ d \ln Z_{q}(\alpha)}{d  \alpha }\, , \nonumber \\
\label{lnZf}  \\
 Z_{q}(\alpha) = Z_{q}(0)
 \exp\left( - \int_{\alpha_0}^{\alpha} d\tau
\frac{\gamma_{q}(\tau)}{\beta(\tau)}
\right) \, .\nonumber
\end{eqnarray}

For definiteness, the bare $Z$-factor $Z(0) $ can be set equal to unity.
It is  obvious that  near fixed point  $\alpha_{*}$, where the $\beta$
function  is zero, only the singular behavior of $1/\beta(\alpha)$ is
important, and the leading contribution comes from the  upper limit  of
integration $\alpha \rightarrow \alpha_{*}$. Therefore, one can
substitute $\gamma_{f}(\tau)$ by a constant,
$\gamma_{f}(\alpha_{*})$.  The $\beta$ function behavior  near the IR
fixed point   is
$\beta(\tau) = -\beta'(\alpha^{*})  (\alpha^{*} - \tau)$,  where
$\beta'(\alpha^{*})$ is  positive  in  the  vicinity of the IR   fixed
point.    After    integration  in (\ref{lnZf})  one gets
\begin{eqnarray}
 Z_{q}(\alpha) \sim
\exp\left(-\frac{\gamma_{q}(\alpha_{*})} {\beta'(\alpha_{*})} \ln
(\alpha_{*}-\alpha)\right) =
\left(\alpha_{*} - \alpha\right)^{ - \gamma_{q}(\alpha_{*})/
\beta'(\alpha_{*})}\, .
\label{theorem1a}
\end{eqnarray}
Since $\gamma_{q}(\alpha_{*}) < 0$ one   immediately
concludes that  at
\begin{eqnarray}
\alpha \rightarrow \alpha_{*}, ~~~~~ Z_{q}(\alpha) \rightarrow
Z_{q}(\alpha_{*}) = 0\;.
\end{eqnarray}

In the general case, $f\neq 0$, the same   result, $Z_q\rightarrow 0$
in the IR fixed point, can be easily obtained by examining  the
following renormalization group invariant combination
\begin{eqnarray}
I \sim \exp\left(-\frac{2\pi}{\alpha(\mu)}\right)
 \mu^{\left( 3 N_c - N_f\right)}
\left(\frac{2\pi}{\alpha(\mu)}\right)^{N_c}  Z_{q}^{-N_{f}}(\mu)
\label{productZ}
\end{eqnarray}
which follows from the  NSVZ $\beta$ function
(\ref{NSVZ}).   The dependence   on the Yukawa coupling  $f$ enters
only through
$Z_{q}(\alpha, f)$.  Now if we reach the IR fixed point $(\alpha_{*},
f_{*})$ at  infinitely large  spatial scale, i.e. at zero $\mu$, to keep
$I$ invariant  we must have
\begin{eqnarray}
 Z_{q}(\alpha_{*}, f_{*}) = 0
\end{eqnarray}
 (and this is the only possibility, because $\alpha_{*}$ is finite).
The condition (\ref{curve1}) implies then that
   $Z_M(\alpha_*, f_*) \rightarrow \infty$
 provided that  $f_{*}$ is finite.

 The  conformal point  for both coupling constant   is defined   now
as the point of the  intersection of curve (\ref{curve1})  with the
curve
\begin{equation}
N_f \gamma_{q}(\alpha_*, f_*)  +(3N_c-N_f) = 0\, .
\label{curve2}
\end{equation}

 In the general  case we do not   know how to find this intersection
point   which  is crucial for the existence of the conformal window.
The consideration summarized in Sec. 2 assumes that the intersection
does exist. We can {\em prove} that it  exists in the weak coupling
regime. In this case it is not difficult to obtain  the  whole phase
portrait of the RG flow for two couplings --
$\alpha(\mu)$ and
$f^{2}(\mu)$ (when both of them  are small).  To this end the  one- loop
results for  $\gamma_{q}(\alpha, f)$ and  $\gamma_{M}(\alpha, f)$  can
be used. It is clear  that the one-loop contribution to
$\gamma_{M}(\alpha, f)$ does not  depend on $\alpha$, because $M$ is a
singlet with  respect to  $SU(N_c)$. It is very easy to calculate this
anomalous dimension; the result is
\begin{eqnarray}
\gamma_{M}(\alpha, f) = \frac{f^{2}}{8 \pi^2} N_c + o(f^2, \alpha)
\end{eqnarray}
 It is also clear that the same formula (with the substitution
$N_c\rightarrow N_f$) will describe  the Yukawa coupling  contribution
to the
 anomalous dimension of the quark superfield.  Combining   with
(\ref{gamma}) we get
\begin{eqnarray}
\gamma_{q}(\alpha, f) =
 -  \frac{\alpha}{\pi}\frac{N_{c}^2 - 1}{2 N_c} +
\frac{f^{2}}{8 \pi^2} N_f
 +  o(f^2, \alpha)\, .
\label{440}
\end{eqnarray}

Two different signs appearing in Eq. (\ref{440}) are perfectly
transparent. The Yukawa contributions to anomalous  dimensions  must be
positive. In pure Yukawa theory one can  not get asymptotic freedom  and
the $\beta$ function  must be positive,  which means positivity for  any
anomalous  dimensions at zero $\alpha$. On the other hand, the  gauge
coupling contribution must be negative. Indeed, the same $Z$ factor
renormalization is responsible for the running of mass, which is  always
asymptotically free in gauge theories.

 At small $f^2$ and $\alpha$ the following  $RG$ equations take place
\begin{eqnarray}
\frac{d\alpha}{d\ln \mu} & =&  -\frac{\alpha^2} {2\pi\left(1-
(N_c\alpha /2\pi)\right)} \left[3N_c -N_f +N_{f}\left(N_f
\frac{f^2}{8\pi^2} -
\frac{N_{c}^2 - 1}{2 N_c}\frac{\alpha}{\pi}
\right)\right] \, ,
   \nonumber \\
\frac{df^2}{d\ln \mu} & =& f^2~\left[\frac{f^{2}}{8\pi^2} (N_c + 2N_f)
- \frac{\alpha}{\pi}\frac{N_c^2-1}{2N_c}\right]\, .
\end{eqnarray}
The corresponding phase portrait is given on Fig. 1. It
is easy
 to see that for $0<3N_c-N_f \ll (N_c,~N_f)$, one has the  infrared
 attractive fixed point
\begin{eqnarray}
\frac{\alpha_*}{\pi} =
\frac{3N_c-N_f}{N_f}\cdot\frac{2N_c}{N_c^2-1}\cdot
\frac{N_c+2N_f}{N_c + N_f} \, ,\nonumber \\
\frac{f^2_*}{8\pi} =
\frac{3N_c-N_f}{N_f}~(N_c+2N_f)\, .
\label{newfixedpoint}
\end{eqnarray}

 Notice that the  old fixed point (i.e. the one in the  theory with the
zero Yukawa coupling)  becomes unstable.

When $3N_c-N_f$ becomes  larger  we must go beyond the one-loop
approximation for  $\gamma$'s to find the IR fixed point which is
defined as the  intersection point for two curves (\ref{curve1}) and
(\ref{curve2}). Although the existence of the IR fixed point is proven
only near the right edge of the window it is natural to expect that
qualitatively the same picture holds in the whole window, up to $N_f  =
(3/2) N_c$. Logically one  cannot exclude, however,  that   at large
$3N_c-N_f$ these   two curves   do not intersect  each other at all --
the RG flow in this case will lead us towards  the strong coupling
region  where in any case we must take into account the effects of
the  pole at   $\alpha_p  = 2\pi/N_c$ in the gauge coupling $\beta$
function.

Note that there exists a domain in the $\{\alpha\, ,\, f^2\}$ plane
where the theory  is asymptotically free  with respect to  both
constants $\alpha$  and $f^2$. In this domain the argument of Sec. 3  is
applicable literally. Outside this domain generically we find ourselves
in the situation with the Landau pole for $f^2$, so that $f^2$ becomes
large in the limit of vanishing distances. Even in this regime,
however, there exists a domain of intermediate distances where $f^2$ is
small (see Fig. 1), so that the inequivalence demonstrated in Sec. 3
will take place in this intermediate domain for sure.

 Summarizing, we have demonstrated that Seiberg's  IR fixed point in
the theory without the Yukawa coupling is unstable; the theory evolves
towards a new  stable IR fixed  point (given by (\ref{newfixedpoint})
in the weak coupling domain). All   quark $Z_q$ factors go to zero near
this fixed point.  At the same time the meson $Z_M$ factor diverges
near this fixed point as $1/Z_q^2$.

 Let us also mention the possibility of  adding  the bare mass  terms
like $m_0  \bar{q}q$ in the theory with the  conformal point.   Because
$Z_{q} \rightarrow 0$  even the infinitesimally small  $m_0$ will  lead
to a large renormalized mass $m = m_0/Z_q$,   and the RG flow,   instead
of approaching the IR conformal point, will  be diverted  to the  strong
coupling region.  It is very important to take into  account the effect
of the vanishing  $Z$ factors when considering the mass   deformations,
for example, in the case of a small explicit SUSY breaking.  At the
same time the bare mass term for the meson field $M$  will  go to zero
near the IR fixed point.  If   one adds both mass and interaction
terms for the  $M$ field,
\begin{eqnarray}
 \int d^4 x d^2 \theta \left(m_0^2 M + \lambda_0 M^3\right)\, ,
\end{eqnarray}
 the renormalized values for $m$ and $\lambda$ are
\begin{eqnarray} m(\mu) = m_{0}/\sqrt{ Z_{M}(\mu)}, ~~ \lambda(\mu)
 = \lambda_0/ Z_{M}^{3/2}(\mu)\, .
\end{eqnarray} They tend  to zero at the fixed point, and at the same
time  the  vacuum expectation value of the renormalized scalar field
$M_R =  M/Z_M^{1/2}$  goes to infinity
\begin{eqnarray}
\langle M^2\rangle \sim  Z_{M}(\mu) (m_0^2/\lambda_0)\, ,
\end{eqnarray}
as well as the quark mass due to the Yukawa coupling
$f_0~M\bar{q}q$
 which will diverge as $m \sim  (m_{0}/\lambda^{1/2})(f_0/Z_q(\mu))$,
 i.e. in the same way as in the  case of the bare quark mass.

\section{Conclusions}

The infrared duality  between two different $N=1$ theories  advocated in
Ref. \cite{S2} seems to be a very promising direction in the studies of
the nonperturbative gauge field dynamics.  We have shown that  the
standard 't Hooft consistency conditions in the external backgrounds,
crucial in  establishing the duality, receive no contributions of the
higher order in $\alpha$ and
$f$ although the higher order corrections enter in the definition of
the conserved $R$ current. In particular, these higher order
corrections are responsible for the fact that the naive (one-loop)
$R$ current defined in the UV flows to $R^0$ residing in the same
supermultiplet with the energy-momentum tensor in IR. Due to specific
holomorphy properties, the corrections in the $R$ current cancel those
appearing in the anomalous 't Hooft triangles, so that the net result
of Ref. \cite{S2} remains intact. The one-loop  nature of the 't Hooft
consistency conditions in the supersymmetric theories  with the matter
fields and a conserved $R$ current is the most  important practical
lesson of our analysis.

The requirement of  the conformal symmetry in the infrared limit in the
magnetic theory yields a constraint on the anomalous dimension of the
$M$ field. If this constraint is satisfied the magnetic theory matches
the electric one in the infrared limit. It  is important that the
conformal symmetry must apply to all
 interactions, including the Yukawa interaction in the magnetic  theory.
 It is also amusing, that only in the stable IR fixed point the
 matching will take place - as we saw there is an unstable IR fixed
 point, where the Yukawa coupling is zero - but there is no matching
 in this point.

Even if  duality is achieved in the infrared limit,  there is  no way
the electric and magnetic theories can coincide identically.

  We established some simple facts about $Z$ factors relevant to the
conformal window and the  infrared attractive fixed points.   A very
general argument   about nullification of the quark $Z$ factors was
presented and possible consequences discussed in brief. Since the old
IR fixed point turns out to be unstable it is important to include the
Yukawa couplings from the very beginning. In the weak coupling  regime
the whole phase portrait in the two-coupling plane is established.

\vspace{1cm}

{\bf Acknowledgments}
 \vskip .2in \noindent We are extremely grateful to N.~Seiberg who
pointed out that the calculation of the matching condition in the
gravitational field in the first version of this paper was erroneous.
Very useful remarks of B. Blok and I.~Pesando are acknowledged. We also
thank S.~Yankielowicz for   discussion of the issues related to  the
short-distance behavior of the electric and magnetic theories and
L.~Koyrakh for  technical assistance with  figures.

    This work was supported in part by DOE under the grant number
DE-FG02-94ER40823, PPARC grant GR/J 21354 and by  Balliol  College,
University of Oxford.

\newpage


\vspace{0.5cm}


\begin{thebibliography}{99}

\bibitem{S2} N. Seiberg, {\it Nucl. Phys.} {\bf B435} (1995) 129.

\bibitem{NSVZ1} V. Novikov, M. Shifman, A. Vainshtein and V. Zakharov,
{\it Nucl. Phys.} {\bf B229} (1983) 407;\\ M. Shifman and A.
Vainshtein, {\it Nucl. Phys.} {\bf B296} (1988) 445. (The
Gell-Mann-Low functions of the type presented in Eq. (\ref{NSVZ}) will
be referred to below as the NSVZ $\beta$ functions. The fact that they
are exact to all orders in $\alpha$ was demonstrated in the first paper
above.   The second  paper actually gives the  proof that the NSVZ
$\beta$ function in $N=1$ theory is exact nonperturbatively although
this was not explicitly stated  there.  In this work the gaugino
condensate
$\langle\lambda\lambda\rangle$ was exactly calculated. Since
$\langle\lambda\lambda\rangle$ is in the same superfield as the anomaly
of the trace of the stress tensor
$\langle\lambda\lambda\rangle$ has zero anomalous dimension; if it is
known exactly the full $\beta$ function immediately follows.)

\bibitem{NSVZ2} V. Novikov, M. Shifman, A. Vainshtein and V. Zakharov,
{\it Nucl. Phys.} {\bf B229} (1983) 381; {\it Phys. Lett.} {\bf B166}
(1986) 329.

\bibitem{S1} N. Seiberg, {\it Phys. Rev.} {\bf D 49} (1994) 6857.

\bibitem{ILS} K. Intriligator, R. Leigh and N. Seiberg, {\it Phys.
Rev.} {\bf D 50}  (1994) 1092.

\bibitem{IS}  K. Intriligator and N. Seiberg, {\it Nucl. Phys.} {\bf
B431} (1994) 551.

\bibitem{Nathan} N. Seiberg, {\em The Power of Duality -- Exact Results
in $D=4$ SUSY Field Theory}, Preprint RU-95-37 [hep-ph/9506077];\\ K.
Intriligator and N. Seiberg, {\em Phases of $N=1$ Supersymmetric Gauge
Theories and Electric-Magnetic Triality}, Preprint RU-95-40
[hep-ph/9506084].

\bibitem{ADS} I. Affleck, M. Dine and N. Seiberg, {\it Nucl. Phys.}
{\bf B241} (1984) 493;  {\bf B256} (1985) 557.

\bibitem{Shif1} M. Shifman and A. Vainshtein, {\it Nucl. Phys.} {\bf
B277} (1986) 456.

\bibitem{Shif2} M. Shifman and A. Vainshtein, {\it Nucl. Phys.} {\bf
B359} (1991) 571.

\bibitem{S3} N. Seiberg, {\it Phys. Lett.} {\bf B318} (1993) 469.

\bibitem{SE} N. Seiberg and E. Witten, {\it Nucl. Phys.} {\bf B426}
(1994) 12; (E) {\bf B430} (1994) 485;  {\bf B431} (1994) 484.

\bibitem{thooft} G. 't Hooft,  in  {\it Recent Developments in Gauge
Theories},
 Eds. G. 't Hooft {\em et al.},  (Plenum Press, New York, 1980).

\bibitem{dolgov}  A. Dolgov and V. Zakharov, {\it Nucl. Phys.}
 {\bf B12}  (1971) 68.

\bibitem{Salam} A. Salam and J. Strathdee, {\it Nucl. Phys.} {\bf B87}
(1975) 85;\\ P. Fayet, {\it Nucl. Phys. } {\bf B90} (1975) 104.

\bibitem{Clark} T.E. Clark, O. Piguet and K. Sibold, {\it Nucl. Phys.}
{\bf B143} (1978) 445;\\ O. Piguet and K. Sibold, {\it Nucl. Phys.}
{\bf B196} (1982) 428, 447.

\bibitem{CPS} T.E. Clark, O. Piguet and K. Sibold, {\it Nucl. Phys.}
 {\bf B159}  (1979) 1; \\{\bf B172} (1980) 201.

\bibitem{Konishi}  K. Konishi, {\it Phys. Lett.} {\bf B135} (1984) 439.

\bibitem{KONS} K. Konishi and K. Shizuya, {\it Nuov. Cim.} {\bf A90}
 (1985) 111.

\bibitem{FZ} S. Ferrara and B. Zumino, {\it Nucl. Phys.} {\bf B87}
(1975) 207.

\bibitem{LS} R. Leigh and M. Strassler, {\em Exactly Marginal Operators
and Duality in Four Dimensional N=1 Supersymmetric Gauge Theory},
 Preprint RU-95-2  [hep-th/9503012].

\bibitem{GRS} M. Grisaru, M. Rocek and W. Siegel, {\it Nucl. Phys.}
 {\bf B159}  (1979) 429.

\end{thebibliography}
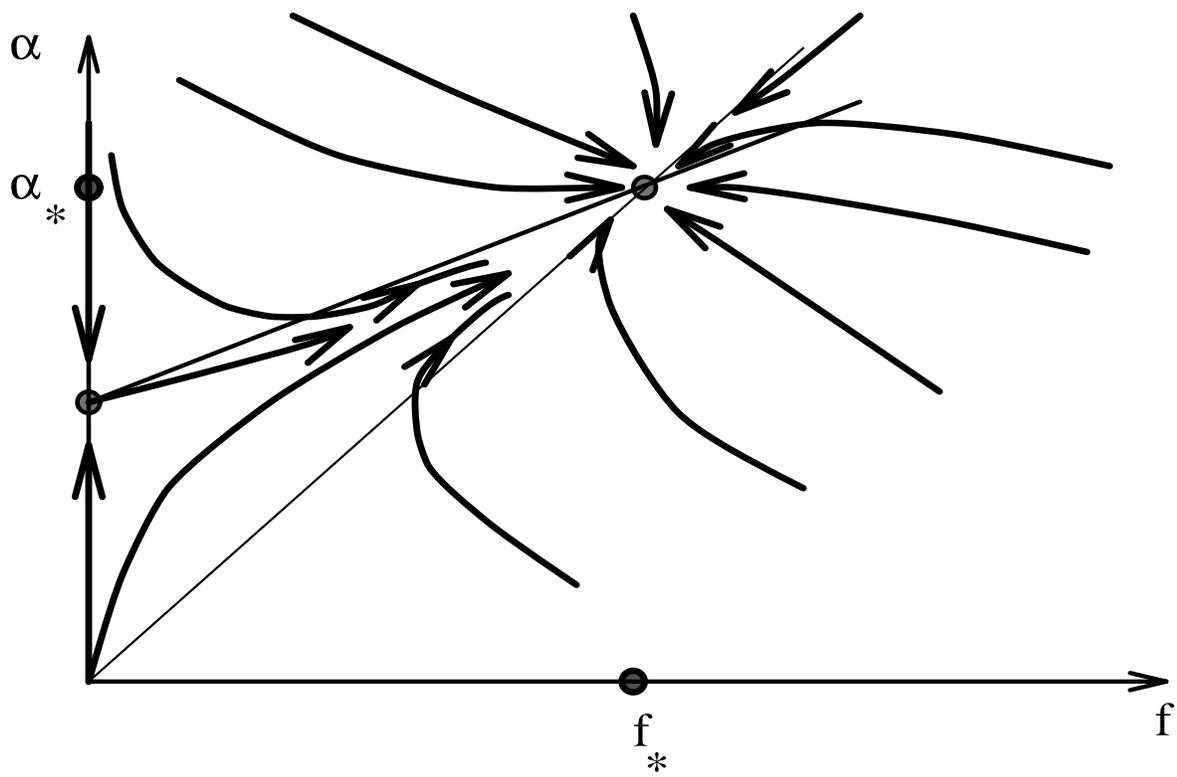
\end{document}